\documentclass[aps,prd,10pt,twocolumn,showpacs,superscriptaddress,nofootinbib]{revtex4-2}
\usepackage{graphicx}
\usepackage{multirow}
\usepackage{xcolor}
\usepackage{enumitem}
\usepackage{booktabs}
\usepackage{slashed}
\usepackage{mathtools}
\usepackage{subfigure}
\usepackage{cleveref}
\crefname{table}{Table}{Tables}
\crefname{equation}{Eq.}{Eqs.}
\crefname{figure}{Fig.}{Figs.}
\crefname{section}{Sec.}{Secs.}

\begin{document}

\title{Near-threshold scattering of proton and Omega baryon and possible bound states }
\author{Yu-Jie Feng}\email{fengyj@ihep.ac.cn}
\affiliation{Institute of High Energy Physics, Chinese Academy of Sciences, Beijing 100049, P.R. China}
\affiliation{University of Chinese Academy of Sciences, Beijing 100049, P.R. China}
\author{Qian Wang}\email{qianwang@m.scnu.edu.cn}
\affiliation{Guangdong Provincial Key Laboratory of Nuclear Science, Institute of Quantum Matter, South China Normal University, Guangzhou 510006, P.R. China}
\affiliation{Guangdong-Hong Kong Joint Laboratory of Quantum Matter, Southern Nuclear Science Computing Center, South China Normal University, Guangzhou 510006, P.R. China}
\affiliation{Southern Center for Nuclear-Science Theory (SCNT), Institute of Modern Physics, Chinese Academy of Sciences, Huizhou 516000, Guangdong Province, P.R. China}
\author{Qiang Zhao}\email{zhaoq@ihep.ac.cn}
\affiliation{Institute of High Energy Physics, Chinese Academy of Sciences, Beijing 100049, P.R. China}
\affiliation{University of Chinese Academy of Sciences, Beijing 100049, P.R. China}
\affiliation{Center for High Energy Physics, Henan Academy of Sciences, Zhengzhou 450046, P.R. China}

\begin{abstract}
    We study the near-threshold scattering and bound-state structure of the $N\Omega$ system by solving the Lippmann–Schwinger (L–S) equation within the framework of the meson exchange model and the Pomeron exchange model. The numerical results indicate that after incorporating the Pomeron exchange mechanism, the observables of the ${^5}S{_2}$ channel, such as the binding energy, scattering length, and effective range, agree better with the experimental measurements. In addition, The Pomeron exchange can provide an extra attractive interaction to make the hadronic state more compact. We also predict the scattering behavior of the ${^3}S{_1}$ channel and confirm that a weak quasi-bound state exists in this channel. Future experimental measurements on the ${^3}S{_1}$ channel will provide an important criterion for verifying the dynamic role played by the Pomeron exchange mechanism within the $N\Omega$ system.
\end{abstract}
\maketitle

\section{Introduction}
Hyperon–nucleon (YN) interaction plays a crucial role in understanding hypernuclear structure and resolving the hyperon puzzle in neutron stars~\cite{Chatterjee:2015pua}. In contrast to nucleon–nucleon (NN) scattering, which possesses high-statistics experimental datasets, YN scattering data remain sparse and limited in precision. This scarcity and low accuracy introduce considerable uncertainties into theoretical descriptions of YN interactions. Beyond direct scattering measurements, observations of possible dibaryon resonances or bound states can likewise provide critical insights and constraints on the YN strong interaction near the threshold. 

Historically, significant progress has been achieved in the study of dibaryon states. The deuteron, the only confirmed bound state to date, demonstrates the critical contribution of tensor forces in NN interactions~\cite{PhysRev.39.164}. Another notable candidate is the H-particle, which carries the same quantum numbers as the S-wave $\Lambda \Lambda$ state and was first predicted by Jaffe~\cite{Jaffe:1976yi} within the bag model. Despite several decades of experimental effort, definitive evidence confirming the existence of the H-dibaryon remains elusive. Moreover, the WASA-at-COSY Collaboration reported the resonance $ d^{*}(2380) $ with quantum numbers $I(J^{P})=0(3^{+})$  in the reacation $ np\to np \pi^0 \pi^0 $~\cite{WASA-at-COSY:2014qkg}, which is hypothesized to correspond to a $ \Delta \Delta $ bound state. An up-to-date overview of $d^*(2380)$ can be found in Ref.~\cite{Dong:2023xdi}.

In recent years, the S-wave proton-Omega ($ p\Omega $) system (also denoted as $ N\Omega $ when the Coulomb interactions are neglected) has emerged as a promising candidate for a bound or quasi-bound state~\cite{STAR:2018uho,Chen:2021hxs,Goldman:1987ma,Haidenbauer:2017sws,Huang:2015yza,HALQCD:2018qyu,Oka:1988yq,Sekihara:2018tsb,Yan:2024aap,Zhang:2025lfn,Li:1999bc,Pang:2003ty,Chen:2011zzb,Sekihara:2023ihc,Dai:2007gc}. For the S-wave $ p\Omega $ dibaryon, there are two possible channels, $ {^3}S{_1} $ and $ {^5}S{_2} $, where $ {^{2S+1}}L{_J} $ denotes the quantum state with spin $S$, relative orbital angular momentum $L$ and total angular momentum $J$. Based on the (2+1) flavor lattice QCD, the HALQCD Collaboration studied the $ N\Omega $ system in the S-wave and spin-2 channel $ ({^5}S{_2}) $ near the physical quark masses. In particular, their numerical result indicates that the $ N\Omega ({^5}S{_2}) $ potential is attractive at all distances, which corresponds to a small $\left | r_{\mathrm{eff}} /a_{0}  \right |$ value,  implying proximity to the unitary limit and  supporting a quasi-bound state with a binding energy of 1.54 MeV~\cite{HALQCD:2018qyu}. Complementary experimental evidence from heavy-ion collisions further corroborates the formation of the $ p\Omega({^5}S{_2})$ bound state: the STAR Collaboration measured the $p-\Omega$ correlation function at $\sqrt{s}=200~\mathrm{GeV}$ and their data strongly favor the $p-\Omega$ bound state hypothesis~\cite{STAR:2018uho,Zhang:2025lfn}.

The observation of this bound state has stimulated extensive phenomenological investigations into its formation mechanism. In Ref.~\cite{Chen:2021hxs}, the method of QCD sum rules was employed to explore the mass spectra of the $ N\Omega $ system. For the ${^5}S{_2}$ state, they found that a bound state may exist with a binding energy of approximately 21 MeV, whereas the mass of the ${^3}S{_1}$ state lies above the corresponding threshold. In the framework of the quark delocalization color screening model (QDCSM), the existence of a $J^{P}= 2^{+}$ bound state (5 MeV binding energy) and an unbound $J^{P}= 1^{+}$ state are supported~\cite{Yan:2024aap}. Based on flavor symmetry, a meson exchange model demonstrates that conventional meson-exchange alone cannot provide enough attraction to bind the $ N\Omega $ system, requiring an additional strong contact interaction~\cite{Sekihara:2018tsb}.

This work aims to understand the origin of the strong attractive interaction in the $N\Omega$ system. Following the effective Lagrangian approach, we propose that soft gluon exchange generates the long-range attraction and employ the Pomeron exchange model to quantify this dynamic process. By incorporating absorption effects arising from open channels, we further analyze the mechanisms underlying the distinct behaviors of different $N\Omega$ quantum states. 

As follows, we first present the theoretical framework in Sec.~\ref{Sec-framework}, and the numerical results will be given in Sec.~\ref{Sec-results}. Conclusions will be given in Sec.~\ref{Sec-conclusion}.

\section{Theoretical Framework}\label{Sec-framework}
For the proton-Omega interaction, the scattering amplitudes at the tree level encompass contributions from both meson exchange and soft gluon exchange. As the latter one does not have a pole in the positive angular momentum complex plane, the two mechanisms do not overlap. Due to isospin conservation in strong interactions, elastic scattering can only involve light isoscalar neutral mesons. Since the $\phi$ and $\omega$ mesons are ideal mixing, their contributions are suppressed according to the OZI rule. Thus, at the tree level, we only consider the exchange of $\eta$, $\sigma$, and $f_{0}(980)$ mesons. The soft gluon exchange is described within the framework of the Pomeron exchange model, which will be discussed later. Furthermore, proton-Omega can also couple to intermediate states through $K$ meson pair exchange serving as a one-loop correction to the elastic scattering process. This amplitude contains the $f_0(980)$ pole contribution via the $S$-wave $K\bar{K}$ scattering. To avoid double-counting we only consider the non-local $K\bar{K}$ exchange as the one-loop contribution in addition to the one meson exchange amplitudes. There are six channels: $\Lambda \Xi$, $\Sigma \Xi$, $\Lambda \Xi^*$, $\Sigma \Xi^*$, $\Sigma^* \Xi$ and $\Sigma^* \Xi^*$ which can couple to $N\Omega$ system. It  has been found in Ref.~\cite{Sekihara:2018tsb} that only the $\Lambda \Xi$ and $\Sigma \Xi$ channels make considerable contributions. In principle, the $\Lambda \Xi$, $\Sigma \Xi$, $N\Omega$ channels should be considered dynamically. As we focus on the absorption effect from those channels,  only the two lowest open channels, i.e. $\Lambda \Xi$ and $\Sigma \Xi$, are included to the box-loop diagrams to incorporate their absorption effect to the $N\Omega\to N\Omega$ scattering. Therefore, the leading-order scattering amplitude $\mathcal{M}$ includes one-boson-exchange, one-loop correction and Pomeron-exchange, expressed as follows: 
\begin{equation}
    \mathcal{M}=\mathcal{M}^{\mathrm{OBE}}+\mathcal{M}^{\mathrm{Loop}}+\mathcal{M}^{\mathrm{Pom}}.
\end{equation}
The non-relativistic interaction potential $ \mathcal{V}_{fi} $ can be obtained from the scattering amplitude $ \mathcal{M}_{fi} $~\cite{Zhao:2013ffn}:
\begin{equation}
    \mathrm{V}_{\mathrm{fi}}=-\frac{\mathcal{M}_{\mathrm{fi}}}{\sqrt{\prod_{\mathrm{f}}E_{\mathrm{f}}/m_{\mathrm{f}} \prod_{\mathrm{i}}E_{\mathrm{i}}/m_{\mathrm{i}}}}.
\end{equation}
\begin{figure*}[htbp]
  \centering
    \subfigure[Pomeron exchange]{\includegraphics[width=0.3\linewidth]{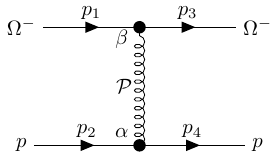}
    \label{subfig:1}}
    \subfigure[One meson exchange]{\includegraphics[width=0.3\linewidth]{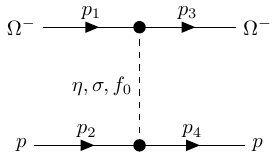}
    \label{subfig:2}}
    \subfigure[Box-loop diagram]{\includegraphics[width=0.3\linewidth]{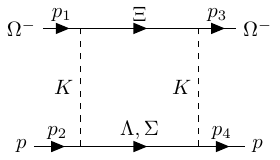}
    \label{subfig:3}}
  \caption{Feynman diagrams of $N\Omega$ scattering process.}
  \label{fig:Feynman diagrams}
\end{figure*}
The full scattering matrix is obtained by introducing the potential kernal into the L-S equation:
\begin{equation}
\begin{split}
    \mathrm{T(E;p^{\prime},p)}&=\mathrm{V(E;p^{\prime},p)}+ \\
    &\mathrm{\int_{0}^{\infty } \frac{\mathrm{d^3}\vec{p}^{\prime\prime} }{(2\pi)^3}} \mathrm{\frac{V(E;p^{\prime},p^{\prime\prime})T(E;p^{\prime\prime},p)}{E-\epsilon -\frac{p^{\prime\prime2}}{2\mu} +i\varepsilon }},
\end{split}
\end{equation}
where $\epsilon$ represents the threshold energy, and $\mu$ is the reduced mass. Meanwhile, $E$ denotes the center-of-mass energy. In this work, the partial-wave potential $V(E;p^{\prime},p)$ is in the $\mathrm{JLS}$ basis, where $\mathrm{J}$, $\mathrm{L}$ and $\mathrm{S}$ denote the total angular momentum, orbital angular momentum and total spin, respectively.
\par
To obtain the  partial wave potential, we use the method given in Refs.~\cite{Jacob:1959at,Sekihara:2018tsb}. For the two-body scattering process $\mathrm{B_{1}(p_{1},\lambda_{1})}+\mathrm{B_{2}(p_{2},\lambda_{2})}\to \mathrm{B_{3}(p_{3},\lambda_{3})}+\mathrm{B_{4}(p_{4},\lambda_{4})}$ in the center-of-mass frame, we can directly obtain the interaction potential in terms of helicity eigenstates as $\mathrm{V(\mathbf{p^{\prime}},\lambda_{3},\lambda_{4},\mathbf{p},\lambda_{1},\lambda_{2})}$. Note that we have chosen the coordinates such that $\mathbf{p}=\mathbf{p_{1}}=-\mathbf{p_{2}}=\mathrm{(0,0,p)}$ and $\mathbf{p^{\prime}}=\mathbf{p_{3}}=-\mathbf{p_{4}}=\mathrm{(p^{\prime}\sin{\theta},0,p^{\prime}\cos{\theta})}$ with the scattering angle $\mathrm{\theta}$. The potential can be projected to the total angular momentum $\mathrm{J}$ as 
        \begin{equation}
        \begin{split}
            \mathrm{V^{J}(p^{\prime},\lambda_{3},\lambda_{4},p,\lambda_{1},\lambda_{2})}&=\frac{1}{2}\int_{-1}^{1}\mathrm{d}\cos{\theta} \ \mathrm{d^{J}_{\lambda_{1}-\lambda_{2}\ \lambda_{3}-\lambda_{4}}(\theta)} \\
            &\times \mathrm{V(\mathbf{p^{\prime}},\lambda_{3},\lambda_{4},\mathbf{p},\lambda_{1},\lambda_{2} )}
        \end{split}
        \end{equation}
where $\mathrm{d^{J}_{m^{\prime}\ m}(\theta)}$ is the Wigner d-matrix function.
\par
The interaction potential $\mathrm{V(E;p^{\prime},p)}$ used for the Lippmann-Schwinger equation can be expressed as $\mathrm{V_{\alpha}(E;p^{\prime},p)}$, where $\mathrm{\alpha}$ is distinguished by the quantum numbers of the initial and final states, specifically $\mathrm{\alpha=(J,P,L^{\prime},S^{\prime},L,S)}$. Here, $\mathrm{J}$ is the total angular momentum, $\mathrm{P}$ represents parity, and $\mathrm{L^{(\prime)}}$ and $\mathrm{S^{(\prime)}}$ correspond to the orbital angular momentum and spin angular momentum of the initial (final) states, respectively. Then, 
$\mathrm{V_{\alpha}}$ can be calculated from
\begin{equation}
\begin{split}
    \mathrm{V_{\alpha}(p^{\prime},p)} &= \sum_{\lambda_{1},\lambda_{2},\lambda_{3},\lambda_{4}} \mathrm{\frac{\sqrt{(2L+1)(2L'+1)}}{2J+1}} \\
    &\mathrm{\times\left \langle j_{3}\ j_{4}\ \lambda_{3}-\lambda_{4}| S^{\prime}\ S^{\prime}_{z} \right \rangle \left \langle L^{\prime}\ S^{\prime}\ 0\ S^{\prime}_{z}| J\ S^{\prime}_{z} \right \rangle }\\
    &\mathrm{\times\left \langle j_{1}\ j_{2}\ \lambda_{1}-\lambda_{2}| S\ S_{z} \right \rangle \left \langle L\ S\ 0\ S_{z}| J\ S_{z} \right \rangle }\\
    &\mathrm{\times V^{J}(p^{\prime},\lambda_{3},\lambda_{4},p,\lambda_{1},\lambda_{2})},
\end{split}
\end{equation}
where $\mathrm{j_{i}}$ is the spin of the baryon $\mathrm{B_{i}}$ , $\mathrm{S_{z}\equiv\lambda_{1}-\lambda_{2}}$ and $\mathrm{S^{\prime}_{z}\equiv\lambda_{3}-\lambda_{4}}$.

\begin{table}[t]
\caption{Pertinent coupling constants for the $\Omega p \to \Omega p$ process~\cite{Wu:2023uva,Sekihara:2018tsb,Rijken:2010zzb}.}
\centering
\begin{tabular}{cccccc}
\toprule
Couplings & $D$ & $F$ & $f_{PBD}$ & $f_{PDD}$ \\
\midrule
Value & 0.795 & 0.465 & 1.8 & 2.09 \\
\midrule
Couplings & $g_{\sigma NN}$ & $g_{f_{0} NN}$ & $g_{\sigma \Omega \Omega}$ & $g_{f_0 \Omega \Omega}$ & \\
\midrule
Value & 8.7 & -4.37 & -0.53 & -13.66 & \\
\bottomrule
\end{tabular}
\label{couplings}
\end{table}

\subsection{Meson exchange}
The effective Lagrangian generating meson exchange are constructed based on SU(3) ChPT and flavor symmetry. The leading-order Lagrangians are given by Refs.~\cite{Wu:2023uva,Sekihara:2018tsb,Rijken:2010zzb}.
    \begin{equation}\label{chiral Lagrangian}
    \begin{split}
        \mathcal{L}_{\mathrm{PBB}}&=-\frac{\mathrm{D}}{\sqrt{2}} \left \langle \bar{\mathrm{B}}\gamma ^{\mu}\gamma_5 \left \{ \partial_{\mu}\Phi ,\mathrm{B} \right \} \right \rangle-\frac{\mathrm{F}}{\sqrt{2}
} \left \langle \bar{\mathrm{B}}\gamma ^{\mu}\gamma_5 \left [ \partial_{\mu}\Phi,\mathrm{B} \right ]  \right \rangle  \\
    \end{split}
    \end{equation}
\begin{align}
          \mathcal{L}_{\mathrm{PBD}}&=-\frac{f_{\mathrm{PBD}}}{m_{\pi}} \left \langle (\bar{T}_{\mu}\cdot \partial^{\mu} \Phi)\mathrm{B}+h.c.  \right \rangle \\
          \mathcal{L}_{\mathrm{PDD}}&=-\frac{f_{\mathrm{PDD}}}{m_{\pi}} \left \langle (\bar{T}^{\mu}\cdot \gamma^{\nu}\gamma^5 T_{\mu})\partial_{\nu} \Phi  \right \rangle\\
          \mathcal{L}_{\sigma NN}&=-g_{\mathrm{\sigma NN}}\sigma (p \bar{p}+n \bar{n})\\
          \mathcal{L}_{f_0 NN}&=-g_{\mathrm{f_0 NN}}f_0 (p \bar{p}+n \bar{n})\\
          \mathcal{L}_{\mathrm{\sigma \Omega \Omega}}&=+g_{\sigma \Omega \Omega} \Omega^- \bar{\Omega}^{+} \sigma\\
          \mathcal{L}_{f_0 \Omega \Omega}&=+g_{f_0 \Omega \Omega} \Omega^- \bar{\Omega}^{+} f_0.
        \end{align}
        In the above Lagrangian, the explicit form of baryon octet are:
        \begin{equation}
           \mathcal{B} =\begin{pmatrix}
  \frac{\Sigma ^{0}}{\sqrt{2} }+\frac{\Lambda  }{\sqrt{6} } & \Sigma ^{+} &p \\
  \Sigma ^{-}& -\frac{\Sigma ^{0}}{\sqrt{2} }+\frac{\Lambda  }{\sqrt{6} } &n \\
  \Xi ^{-}& \Xi ^{0} &-\frac{2}{\sqrt{6} } \Lambda
\end{pmatrix},
       \end{equation}
       and the pseudoscalar meson octet is given by 
       \begin{equation}
          \Phi =\begin{pmatrix}
  \frac{\pi ^{0}}{\sqrt{2} }+\frac{\eta  }{\sqrt{6} } & \pi ^{+} &K^{+} \\
  \pi ^{-}& -\frac{\pi ^{0}}{\sqrt{2} }+\frac{\eta  }{\sqrt{6} } &K^{0} \\
  K^{-}& \bar{K ^{0}} &-\frac{2}{\sqrt{6} } \eta
\end{pmatrix}.
       \end{equation}
In the above Lagrangians, the explicit form of the decuplet baryon totally symmetric tensor field $T$ are assigned as~\cite{Ren:2013oaa,Haidenbauer:2017sws}:
       \begin{equation}
            \begin{aligned}
                &T^{111}=\Delta^{++},\ \ \ T^{112}=\frac{1}{\sqrt{3} }\Delta^{+}, \ T^{113}=\frac{1}{\sqrt{3} }\Sigma^{*+},\\
&T^{122}=\frac{1}{\sqrt{3} }\Delta^{0},\  T^{123}=\frac{1}{\sqrt{6} }\Sigma^{*0},\ T^{133}=\frac{1}{\sqrt{3} }\Xi^{*0},\\
&T^{222}=\Delta^{-},\ \ \ \ \ T^{223}=\frac{1}{\sqrt{3} }\Sigma^{*-},\ T^{233}=\frac{1}{\sqrt{3} }\Xi^{*-},\\
&T^{333}=\Omega^{-}.
            \end{aligned}
        \end{equation}
\par
The pseudoscalar-baryon-baryon coupling constants  have been well determined by fitting experimental data. In the present work, we adopt the values from Ref.~\cite{Sekihara:2018tsb}.
The scalar-nucleon-nucleon coupling constants are calculated using the correlated $\pi\pi$ interaction model~\cite{Wu:2023uva} and the Nijmegen extended-soft-core model~\cite{Rijken:2010zzb}.
The coupling constants $g_{\sigma \Omega \Omega}$ and $g_{f_0 \Omega \Omega}$ can be derived from $\left | g_{\sigma \Delta \Delta} \right | =1.11\left | g_{\sigma N N} \right | $~\cite{Lopes:2022vjx} via SU(3)-flavor symmetry, which will be illustrated in the following.

In fact, the nature of the scalar nonet mesons with masses below 1 GeV remains controversial, and the constraints on the coupling constants depend on the specific theoretical model used to describe scalar mesons.
In the quark model, scalar mesons are conventionally regarded as $^{3}P_{0}$ $q\bar{q}$ states.
However, some studies interpret them as tetraquark states $qq\bar{q}\bar{q}$.
In recent years, the idea that scalar mesons could be dynamically generated resonances from meson-meson interactions~\cite{Hanhart:2007wa,Branz:2007xp,Liang:2016hmr,Molina:2019udw,Dai:2023jix} has gained increasing support.
Accordingly, it is unclear whether it is reliable to impose SU(3) symmetry.
For simplicity, we nevertheless adopt the $q\bar{q}$ scheme and assume the constraints of SU(3) symmetry. According to the decomposition $\mathbf{10}\otimes \mathbf{\bar{10}}=\mathbf{64} \oplus \mathbf{27} \oplus\mathbf{8} \oplus\mathbf{1}$~\cite{deSwart:1963pdg}, we consider the SU(3)-invariant Lagrangian
        \begin{equation}\label{SU(3) Lagrangian}
            \mathcal{L}=g_8\left\langle \bar{T}^{\mu}T_{\mu}\phi_8 \right\rangle + g_1\left\langle \bar{T}^{\mu}T_{\mu}\right\rangle\left\langle\phi_1 \right\rangle,
        \end{equation}
where $\phi_8$ and $\phi_1$ denote the scalar octet and singlet, respectively. And their matrix representation are:
\begin{equation}\label{octetscalar}
\begin{aligned}
\phi_{8} &=\begin{pmatrix}
  \frac{a_{0}  ^{0}}{\sqrt{2} }+\frac{\sigma _{8}  }{\sqrt{6} } & a_{0} ^{+} &\kappa ^{+} \\
  a_{0} ^{-}& -\frac{a_{0} ^{0}}{\sqrt{2} }+\frac{\sigma_{8}  }{\sqrt{6} } &\kappa^{0} \\
  \kappa^{-}& \bar{\kappa ^{0}} &-\frac{2}{\sqrt{6} } \sigma_{8}
\end{pmatrix},\\
       \phi_{1} &=\frac{1}{\sqrt{3} } \begin{pmatrix}
  \sigma_{1} & 0 &0 \\
  0& \sigma_{1}&0 \\
  0& 0 &\sigma_{1}
\end{pmatrix}.
\end{aligned}
\end{equation}
The physical states are mixing of the pure $ \mathrm{SU}(3) $ states
        \begin{equation}
            \binom{f_{0}}{\sigma}=\begin{pmatrix}
        \cos \theta_{s} &-\sin  \theta_{s}\\
        \sin  \theta_{s}&\cos \theta_{s}
        \end{pmatrix} \binom{\sigma_{8}}{\sigma_{1}}
        \end{equation}
with $\theta_{s}=37.5^{\circ}$~\cite{Rijken:2010zzb}. One can define the effective Lagrangian of $\sigma \Delta \Delta$ as
\begin{equation}\label{DDs Lagrangian}
    \mathcal{L}_{\sigma \Delta \Delta}=g_{\sigma \Delta \Delta}\bar{\Delta}\Delta\sigma,
\end{equation}
where $\Delta=(\Delta^{++},\Delta^{+},\Delta^{0},\Delta^{-})^{\mathrm{T}}$. As for singlet coupling $g_1$, it can be estimated on the basis of the constituent quark of $ \sigma $ and $ \Delta $. Specifically, in the $ \left \{ u,d,s \right \} $ basis,
        \begin{equation}
            \begin{aligned}
                \sigma_{1} & = \frac{1}{\sqrt{3} } (u\bar{u}+d\bar{d}+s\bar{s})\\
\sigma_{8} & = \frac{1}{\sqrt{6} } (u\bar{u}+d\bar{d}-2s\bar{s}).
            \end{aligned}
        \end{equation}
Due to the fact that there is no strange constituent quark in the $\Delta$ baryon, we can assume that only the $ u $, $ d $ quark contributes to the interaction of $ \Delta \Delta \sigma $. This estimate gives $g_{\sigma \Delta \Delta_{1}}=\sqrt{2}g_{\sigma \Delta \Delta_{8}} $, or equivalently $ g_{\sigma \Delta \Delta}=\frac{1}{\sqrt{2} }(\mathrm{sin{\theta_s}+cos{\theta_s}})g_{\sigma \Delta \Delta_1} $. Thus, the Lagrangian \cref{SU(3) Lagrangian} is fully determined by $g_{\sigma \Delta \Delta}$. The relative sign between $g_{\sigma NN}$ and $g_{\sigma \Delta \Delta}$ is determined by the quark model and the technical details are discussed in Ref.~\cite{Yalikun:2021dpk}. In a word, one can calculate the interaction current at quark level and at hadronic level, respectively, and then equate them to get the right sign of the coupling constants. If the effective Lagrangian describing the interaction between light quarks and $\sigma$ meson takes the form:
\begin{equation}
    \mathcal{L}_{q}=-g_{qq\sigma}\bar{\psi }_{q} \sigma \psi_{q},
\end{equation}
the $g_{\sigma N N}$ and $g_{qq\sigma}$ have the same sign. Due to a minus sign introduced by the polarization vector in the normalization of the $\Delta$ baryon, the coupling $g_{\sigma \Delta \Delta}$ in \cref{DDs Lagrangian} carries the same sign as $g_{\sigma NN}$. The remaining couplings are obtained via SU(3) symmetry:
         \begin{equation}
            g_{f_0 \Omega \Omega}=-\sqrt{2} g_{\sigma \Delta \Delta},
         \end{equation}
and
         \begin{equation}
            g_{\sigma  \Omega \Omega}=\frac{(-\sqrt{2}\sin{\theta_{s}}+\cos{\theta_{s}})g_{\sigma \Delta \Delta}}{\frac{1}{\sqrt{2}}\sin{\theta_{s}}+\cos{\theta_{s}}}.
         \end{equation}
\par
With the above Lagrangian, one can directly express the interaction amplitude of $N\Omega$ scattering as
\begin{align}
  \mathcal{M}_{\eta}
  &= \frac{2\sqrt{2}(D-3F)f_{\mathrm{PDD}}(M_pM_{\Lambda})F(q^2,\Lambda_m)^2}
          {3f_{\eta}m_{\pi}(q^2-m_{\eta}^2)}
     \nonumber\\
  &\quad\times \bar{u}_{\mu}(p_3)\gamma_5u^{\mu}(p_1)\,
     \bar{u}_N(p_4)\gamma_5u_N(p_1), \\[4pt]
  \mathcal{M}_{\sigma}
  &= \frac{g_{\sigma NN}\,g_{\sigma\Omega\Omega}F(q^2,\Lambda_m)^2}
          {q^2-m_{\sigma}^2}
     \nonumber\\
  &\quad\times \bar{u}_{\mu}(p_3)u^{\mu}(p_1)\,
     \bar{u}_N(p_4)u_N(p_1), \\[4pt]
  \mathcal{M}_{f_0}
  &= \frac{g_{f_0 NN}\,g_{f_0\Omega\Omega}F(q^2,\Lambda_m)^2}
          {q^2-m_{f_0}^2}
     \nonumber\\
  &\quad\times \bar{u}_{\mu}(p_3)u^{\mu}(p_1)\,
     \bar{u}_N(p_4)u_N(p_1), \\[4pt]
  \mathcal{M}_{K}^{\Lambda\Xi}
  &= \frac{-(D+3F)f_{\mathrm{PBD}}(M_p+M_{\Lambda})F(q^2,\Lambda_m)^2}
          {2\sqrt{3}f_k m_{\pi}(q^2-m_k^2)}
     \nonumber\\
  &\quad\times \bar{u}_{\Xi}(p_3)q^{\mu}u_{\mu}(p_1)\,
     \bar{u}_{\Lambda}(p_4)\gamma_5u_{N}(p_2), \\[4pt]
  \mathcal{M}_{K}^{\Sigma\Xi}
  &= \frac{-\sqrt{3}(D-F)f_{\mathrm{PBD}}(M_p+M_{\Sigma})F(q^2,\Lambda_m)^2}
          {2f_k m_{\pi}(q^2-m_k^2)}
     \nonumber\\
  &\quad\times \bar{u}_{\Xi}(p_3)q^{\mu}u_{\mu}(p_1)\,
     \bar{u}_{\Sigma}(p_4)\gamma_5u_{N}(p_2).
\end{align}
where we adopt $f_{\pi}=92.1 ~\mathrm{MeV}$, $f_{k}=1.2f_{\pi}$ and $f_{\eta}=1.3f_{\pi}$ to reflect chiral symmetry breaking in \cref{chiral Lagrangian}. As usual in effective field theory, we introduce the form factor $F(q^2,\Lambda_m)=\frac{\Lambda_m^2}{\Lambda_m^2+\left | q^2 \right |}$ for t-channel meson exchange processes. Here, $\left | q^2 \right |\equiv -q^2$ is used to avoid unphysical singularity in analytic continuation. In the present work, the value of the cut-off $\Lambda_m$ is fixed as $0.8~\mathrm{GeV}$ as a typical scale. The interaction potential of box diagrams can be expressed as non-relativistic integral representation:
\begin{equation}
    \begin{aligned}
        \mathrm{V_{\mathrm{box}}^{\Lambda \Xi}}&=\int_{0}^{\infty } \frac{\mathrm{d^3}\vec{p}^{''} }{(2\pi)^3} \frac{\mathrm{V^{\Lambda\Xi \to N\Omega}(E;p^{\prime},p^{''})V^{N\Omega \to \Lambda\Xi}(E;p^{''},p)}}{\mathrm{E}-\epsilon_{\Lambda\Xi} -\mathrm{\frac{p^{''2}}{2\mu_{\Lambda\Xi}}} +i\varepsilon },\\
        \mathrm{V_{\mathrm{box}}^{\Sigma \Xi}}&=\int_{0}^{\infty } \frac{\mathrm{d^3}\vec{p}^{''} }{(2\pi)^3} \frac{\mathrm{V^{\Sigma\Xi \to N\Omega}(E;p^{\prime},p^{''})V^{N\Omega \to \Sigma\Xi}(E;p^{''},p)}}{\mathrm{E}-\epsilon_{\Sigma\Xi} -\mathrm{\frac{p^{''2}}{2\mu_{\Sigma\Xi}}} +i\varepsilon }.
    \end{aligned}
\end{equation}

\subsection{Pomeron exchange}
The Pomeron exchange model serves as an effective theory describing multi-soft-gluon exchanges in the $t$-channel, and is widely applied to account for the behavior of high-energy hadron-hadron collisions as well as the diffractive pattern of vector meson photoproduction off nucleons~\cite{Donnachie:1984xq,Donnachie:1987pu,Pichowsky:1996jx,Pichowsky:1996tn,Laget:1994ba,Zhao:1999af}. It should be emphasized that the Pomeron exchange model is only valid under the Regge condition $s \gg \left | t \right | $. Thus, it is often applied to high-energy scatterings, in particular, in processes involving only light flavor hadrons such as $pp$ scatterings. For the threshold region, one notices that  the Regge condition $s\gg |t|$ can be approximately fulfilled except that the kinematic region is rather limited.  Taking into account that in those two-body light hadron systems the  single-pion or single-kaon exchange if allowed is often the dominant $t$-channel transition, the Pomeron exchange contribution is generally neglected near threshold. However, in processes where the  single-pion or single-kaon exchange is forbidden at leading order, the Pomeron exchange mechanism may become a leading term in the hadron interactions and should not be neglected~\cite{Gong:2020bmg,Gong:2022hgd}. In the case of the $N\Omega$ scattering, we are interested in examining whether there exist some space for the Pomeron exchange since here the leading single-pion and kaon exchange are forbidden.

In the Pomeron exchange model, the amplitude for $\mathrm{A+B\to A+B}$ scattering at high $s$ becomes
        \begin{equation}
        \begin{split}
            \mathrm{T(s,t)}&=\left [ \beta_{\mathrm{q_{A}}}n_{\mathrm{A}}F_{\mathrm{A}}(t) \right ]\left [ \beta_{\mathrm{q_{B}}}n_{\mathrm{B}}F_{\mathrm{B}}(t) \right ]  \\
            &\times \frac{e^{-i\frac{\pi}{2} (\alpha_{\mathrm{p}}(t)-1)}}{2\sin{(\frac{\pi}{2} \alpha_{\mathrm{p}}(t))}}(\alpha_{\mathrm{1,p}}s)^{\alpha_{\mathrm{p}}(t)},
        \end{split}
        \end{equation}
where $n_{A}$ is the number of constituent quarks in hadron A and $\beta_{\mathrm{q_{A}}}$ is the quark-Pomeron coupling constant. For $p+\Omega^{-} \to p+\Omega^{-}$, one has $n_{\mathrm{p}}=n_{\Omega}=3$ and $\beta_{\mathrm{q_{p}}}=\beta_{\mathrm{u}}=\beta_{\mathrm{d}}=2.07\ \mathrm{GeV^{-1}},\beta_{\mathrm{q_{\Omega}}}=\beta_{s}=1.38\ \mathrm{GeV^{-1}}$~\cite{Lee:2022ymp}. The Regge trajectory of Pomeron is described as 
        \begin{equation}
            \alpha_{\mathrm{p}}(t)=\alpha_{0,p}+\alpha_{\mathrm{1,p}}t=1.08+0.25t \ ,
        \end{equation}
where $t$ is in unit of $\mathrm{GeV^{2}}$. As an effective model for the multi-soft-gluon exchange, the Pomeron exchange manifests as the exchange of isoscalar photons with quantum numbers $J^{\mathrm{PC}}=1^{-+}$. Therefore, the Pomeron-nucleon vertex is described by $F_{\alpha}(t)=\gamma_{\alpha}f_{\mathrm{N}}(t)$ where $f_{\mathrm{N}}(t)$ is the form factor of nucleon. Actually, this coupling involves two form factors $f_{1}(t)$ and $f_{2}(t)$, the latter corresponding to helicity flip of the nucleon. However,  experimental evidence indicates that the Pomeron shows very small helicity-flipping contribution. Thus, we can neglect this helicity-flipping term and the form factor can be written as~\cite{Donnachie:1984xq}
        \begin{equation}
            f_{1}(t)=\frac{4m_{\mathrm{N}}^{2}-2.8t}{(4m_{\mathrm{N}}^{2}-t)(1-\frac{t}{0.71})^{2}}.
        \end{equation}
\par
The same strategy can be employed to investigate the Lorentz structure of the Pomeron-Omega vertex. Noting that the electromagnetic current $J^{\mu}$ associated with the $\gamma^{*} \Omega^{-} \bar{\Omega}^{+}$ vertex can be written in the Lorentz-invariant and gauge-invariance form as~\cite{Nozawa:1990gt,Alexandrou:2009hs,Ramalho:2020laj}
        \begin{equation}
            \begin{aligned}
                J^{\mu}=&-\left [ F_{1}^{*}(Q^{2})g^{\alpha \beta}+F_{3}^{*}(Q^{2})\frac{q^{\alpha}q^{\beta}}{4M^{2}} \right ]\gamma^{\mu}\\
 &-\left [F_{2}^{*}(Q^{2})g^{\alpha \beta}+F_{4}^{*}(Q^{2})\frac{q^{\alpha}q^{\beta}}{4M^{2}}  \right ]\frac{i\sigma ^{\mu \nu}}{2M},
            \end{aligned}
        \end{equation}
in elementary charge units ($e$). The value of the charge ($e_{\Omega}=-1$) indicates that $F_{1}^{*}(0)=G_{\mathrm{E0}}(0)=-1$. Since the Pomeron is equivalent to an isoscalar photon, the Lorentz structure of the Pomeron-$\Omega$ vertex should be similar to that of photon-$\Omega$ vertex, with the helicity-flipping term neglected. The multipole form factors, $G_{\mathrm{E0}}$, $G_{\mathrm{M1}}$, $G_{\mathrm{E2}}$, and $G_{\mathrm{M3}}$ can be expressed as linear combinations of the elementary form factors $F_{i}^{*}$, that is~\cite{Alexandrou:2009hs,Ramalho:2020laj}
        \begin{equation}
            \begin{aligned}
                G_{\mathrm{E0}}&=(1+\frac{2}{3}\tau)(F_{1}^{*}-\tau F_{2}^{*})-\frac{1}{3}\tau (1+\tau)(F_{3}^{*}-\tau F_{4}^{*}) \\
                G_{\mathrm{E2}}&=(F_{1}^{*}-\tau F_{2}^{*})-\frac{1}{2}(1+\tau)(F_{3}^{*}-\tau F_{4}^{*}) \\
                G_{\mathrm{M1}}&=(1+\frac{4}{5}\tau)(F_{1}^{*}+F_{2}^{*})-\frac{1}{2}(1+\tau)(F_{3}^{*}-\tau F_{4}^{*}) \\
                G_{\mathrm{M3}}&=(F_{1}^{*}+F_{2}^{*})-\frac{1}{2}(1+\tau)(F_{3}^{*}+F_{4}^{*})
            \end{aligned}
        \end{equation}
with $\tau=-\frac{q^2}{4M_{\Omega}^{2}}$. For large $Q^{2}$, one can estimate the falloff of the form factors~\cite{Ramalho:2020laj}:
        \begin{equation}
            F_{1}^{*}(Q^{2})\propto \mathcal{O}(\frac{1}{Q^{4}} ),\ \ F_{3}^{*}(Q^{2})\propto \mathcal{O}(\frac{1}{Q^{6}} )
        \end{equation}
with asymptotic value $G_{\mathrm{M1}}(0)=-3.60\pm 0.09$ ~\cite{ParticleDataGroup:2020ssz} and $G_{\mathrm{E2}}(0)=0.680\pm0.012$~\cite{Ramalho:2010rr} at $Q^{2}=0$. We can always express the form factors in the following manner:
        \begin{equation}\label{FF-1}
        \begin{aligned}
            F_{1}^{*}(t)&=-\frac{1}{(1-\frac{t}{\Lambda_{p}^2})^2},\\
            F_{3}^{*}(Q^2)&=\frac{F_{3}^{*}(0)}{(1-\frac{t}{\Lambda_{p}^2})^3}\approx-\frac{3.36}{(1-\frac{t}{\Lambda_{p}^2})^3},
        \end{aligned}
        \end{equation}
where $\Lambda_{p}$ is the cut-off as a free variable. The Pomeron exchange potential of $p-\Omega$ scattering can be written as
        \begin{equation}
        \begin{split}
           \mathcal{M}^{\mathrm{Pom}}&=i\bar{u}^{\mu}(\vec{p}^{\prime},\lambda_{3})\left [ 3\beta_{s}(F_{1}^{*}(t)g_{\mu\nu}+F_{3}^{*}(t)\frac{q_{\mu}q_{\nu}}{4M_{\Omega}^2} ) \right ]\\
           &\times \gamma_{\beta} u^{\nu}(\vec{p},\lambda_{1})\bar{u}(-\vec{p}^{\prime},\lambda_{4})\left [ 3\beta_{\mathrm{u/d}}f_{1}(t) \right ]\\
           &\times \gamma^{\beta}u(-\vec{p},\lambda_{2})G_{p}(t)
        \end{split}
        \end{equation}
with $q=p_{3}-p_{1}$ is the momentum of Pomeron, and $G_{p}(t)$ is the propagator of Pomeron in the Regge phenomenology and can be written as $G_{p}(t)=(-i\alpha_{1,p}s)^{\alpha_{p}(t)-1}$~\cite{Pichowsky:1996jx,Zhang:2024dkm}.

\section{Numerical Results And Discussions}\label{Sec-results}

\begin{table}[t]
\caption{The scattering length for the $\Omega p \to \Omega p$ process at different value of $\Lambda_p$ (in units of GeV). Spin ave denotes the statistical weight average over all possible spin multiplets of the particle pair, while Quintet refers to an independent fit to the quintet channel with total spin $S=2$ in the Lednicky–Lyuboshitz correlation function analysis in heavy-ion collisions. HALQCD denotes the lattice QCD calculation result for the ${^5}S{_2}$ channel. All data are taken from Refs.~\cite{HALQCD:2018qyu,Zhang:2025lfn}.}
\centering
\begin{tabular}{ccccc}
\toprule
 (GeV)& $\Lambda_{p}=0 $ & $\Lambda_{p}=0.1$ & $\Lambda_{p}=0.15$ & $\Lambda_{p}=0.2 $ \\
\midrule
${^5}S{_2}$ & $6.99-1.06i$ & $5.76-2.39i$ & $4.87-2.49i$ & $4.11-2.47i$\\
${^3}S{_1}$ & $109-6.09i$ & $6.18-15.86i$ & $3.79-9.05i$ & $2.90-6.41i$\\
\midrule
 & $\Lambda_{p}=0.25$ & Spin ave & Quintet & HALQCD \\
\midrule
${^5}S{_2}$ & $2.11-5.19i$ & $4.89_{-0.54}^{+0.66}$ & $4.27_{-0.43}^{+0.68}$ & $5.30_{-0.01}^{+0.16}$ \\
${^3}S{_1}$ & $2.38-5.05i$ & $\mathrm{none}$ & $\mathrm{none}$ & $\mathrm{none}$ \\
\bottomrule
\end{tabular}
\label{Scattering Length}
\end{table}

\begin{table}[tbp]
\caption{The effective range for the $\Omega p \to \Omega p$ process at different value of $\Lambda_p$ (in units of GeV).}
\centering
\begin{tabular}{ccccc}
\toprule
 (GeV)& $\Lambda_{p}=0$ & $\Lambda_{p}=0.1$ & $\Lambda_{p}=0.15$ & $\Lambda_{p}=0.2$ \\
\midrule
${^5}S{_2}$ & $0.99+0.22i$ & $1.01+0.08i$ & $1.18+0.25i$ & $1.26+0.30i$\\
${^3}S{_1}$ & $1.07+0.002i$ & $2.67-1.24i$ & $3.52-2.30i$ & $4.04-3.63i$\\
\midrule
& $\Lambda_{p}=0.25$ & Spin ave & Quintet & HALQCD \\
\midrule
${^5}S{_2}$ & $1.28+0.29i$ & $2.32_{-0.52}^{+0.44}$ & $1.53_{-0.66}^{+0.54}$ & $1.26_{-0.01}^{+0.02}$ \\
${^3}S{_1}$ & $4.37-5.04i$ & $\mathrm{none}$ & $\mathrm{none}$ & $\mathrm{none}$ \\
\bottomrule
\end{tabular}
\label{Effective Range}
\end{table}

\begin{table}[t]
\caption{Binding energy for the $\Omega p \to \Omega p$  process.}
\centering
\begin{tabular}{ccccc}
\toprule
 (GeV)& $\Lambda_{p}=0$ & $\Lambda_{p}=0.1$ & $\Lambda_{p}=0.15$ & $\Lambda_{p}=0.2$ \\
\midrule
$\mathrm{{^5}S{_2}(MeV)}$ & $1.61+0.37i$ & $1.68+0.96i$ & $1.73+1.45i$ & $1.77+1.94i$\\
$\mathrm{{^3}S{_1}(MeV)}$ & $0.09+0.005i$ & $0.18+0.57i$ & $0.21+1.04i$ & $0.24+1.48i$\\
\midrule
 & $\Lambda_{p}=0.25$ & Spin ave & Quintet & HALQCD \\
\midrule
$\mathrm{{^5}S{_2}(MeV)}$ & $1.79+2.40i$ & $1.51_{-0.56}^{+1.12}$ & $1.55_{-0.50}^{+1.44}$ & $1.54_{-0.10}^{+0.04}$ \\
$\mathrm{{^3}S{_1}(MeV)}$ & $0.26+1.89i$ & $\mathrm{none}$ & $\mathrm{none}$ & $\mathrm{none}$ \\
\bottomrule
\end{tabular}
\label{Binding Energy}
\end{table}
In this section, we discuss the numerical results for the $N\Omega$ scattering near the threshold within our framework. We mainly focus on the behavior of the interaction potential and related observables. We first briefly describe how to extract observable physical quantities from the full scattering amplitude $\mathrm{T(E;k,k)}$ obtained by solving the L-S equation. In the present notation, the $N\Omega$ scattering amplitude in the nonrelativistic quantum mechanics is expressed as
        \begin{equation}
            f_{s}(\mathrm{k})=-\frac{\mu}{2\pi}\mathrm{T(E;k,k)}.
        \end{equation}
For the region where momentum $\mathrm{k}$ is close to zero, the inverse of the scattering amplitude $f_{s}^{-1}(\mathrm{k})$ can be expanded as a in terms of $\mathrm{k}$ as
        \begin{equation}
            f_{s}(\mathrm{k})^{-1}=-\frac{1}{a}-i\mathrm{k}+\frac{1}{2}r_{\mathrm{eff}}\mathrm{k^{2}}+\mathcal{O}(\mathrm{k^{4}}).
        \end{equation}
\begin{figure}[htbp!]   
    \centering
    \subfigure[]{\includegraphics[width=0.48\textwidth]{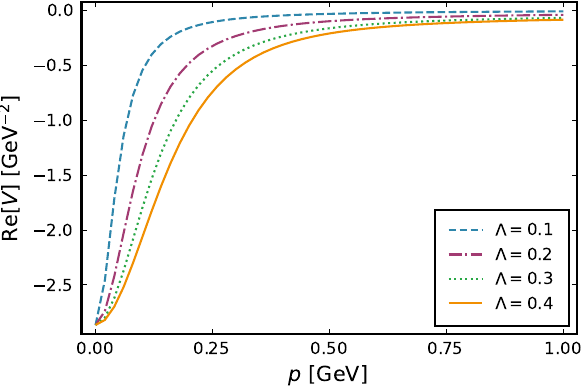}
    \label{subfig1}}
    
    \subfigure[]{\includegraphics[width=0.48\textwidth]{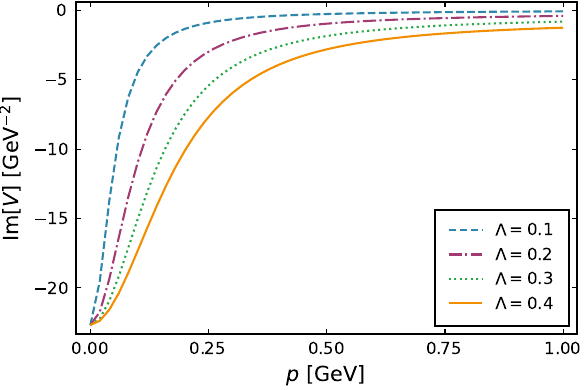}
    \label{subfig2}}
  \caption{Behavior of Pomeron-exchange interaction potential under different values of $\Lambda_p$. The horizontal axis $p$ denotes the magnitude of the incident particle momentum in the center-of-mass frame.}
  \label{fig2}
\end{figure}

Then, we can obtain low-energy parameters from the behavior of the scattering amplitude at the threshold:
        \begin{equation}
            \begin{aligned}
                a&=-f_{s}(\mathrm{k=0})\\
                r_{\mathrm{eff}}&=2\left [ \frac{\mathrm{d} f_{s}^{-1}(\mathrm{k})}{\mathrm{d} \mathrm{k^{2}}}  \right ]_{\mathrm{k=0}}
            \end{aligned}
        \end{equation}
Regarding the notation for the scattering length, it is conventionally accepted that $a>0$ indicates either a repulsive interaction or the presence of a bound state below the threshold; whereas $a<0$ signifies that attractive interactions are present but are not strong enough to form a bound state. The binding energy in our notation can be evaluated as
        \begin{equation}
            \mathrm{B}=\mathrm{E}_{\mathrm{pole}}-\epsilon,
        \end{equation}
where $E_{\mathrm{pole}}$ is the position of the pole. The Weinberg criterion can be used to characterise the compositeness 
\begin{equation}
    \bar{X} =\sqrt{\frac{1}{1+\left | 2r/a \right | } }
\end{equation}
of bound, virtual and resonance states~\cite{Baru:2021ldu}.

\begin{figure}[t]  
  \centering          
  \includegraphics[width=\linewidth]{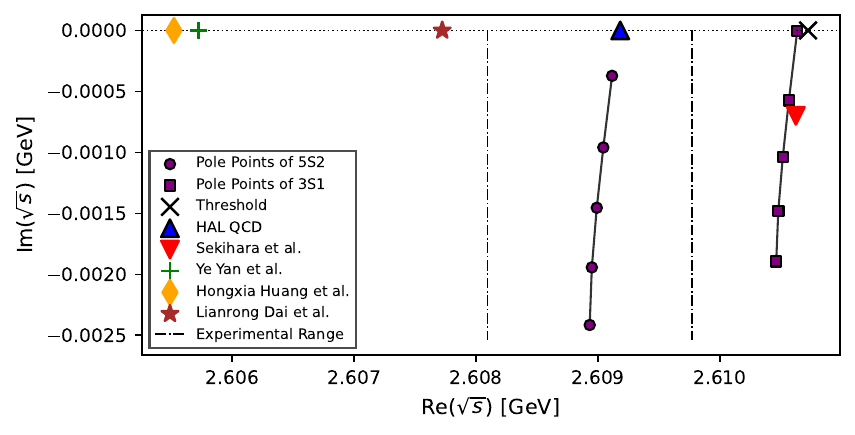} 
  \caption{Trajectory of poles in the first Riemann sheet. The circles and squares denote the pole locations for $N\Omega({^5}S{_2})$ and $N\Omega({^3}S{_1})$, respectively. All poles from top to bottom correspond to the cutoff values $\Lambda_p=0,0.1,0.15,0.2,0.25$ GeV. For comparison, theoretical predictions for ${^5}S{_2}$ are shown from the HALQCD~\cite{HALQCD:2018qyu} (blue triangle up), Sekihare et al.~\cite{Sekihara:2018tsb} (red triangle down), Yan et al.~\cite{Yan:2024aap} (green plus), Huang et al.~\cite{Huang:2015yza} (orange diamond), and Dai et al.~\cite{Dai:2007gc} (brown star). The region enclosed by the two dot-dashed lines indicates the range allowed by experimental uncertainties~\cite{Zhang:2025lfn}.}
  \label{fig6} 
\end{figure}

\begin{figure}[t]  
  \centering          
  \includegraphics[width=\linewidth]{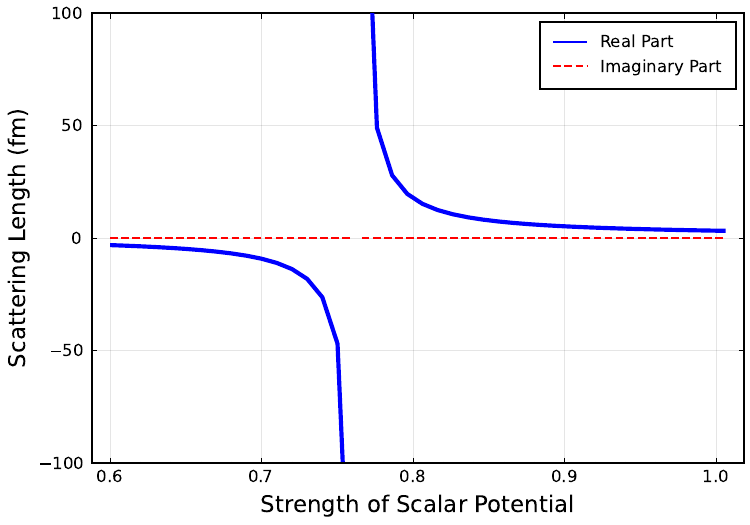} 
  \caption{Dependence of the scattering length on the strength of the scalar potential associated with $f_{0}(980)$ meson exchange, where the interaction strength is controlled by varying the corresponding coupling constant.}
  \label{fig7} 
\end{figure}

\begin{figure*}[t]   
    \centering
    \subfigure[]{\includegraphics[width=0.48\textwidth]{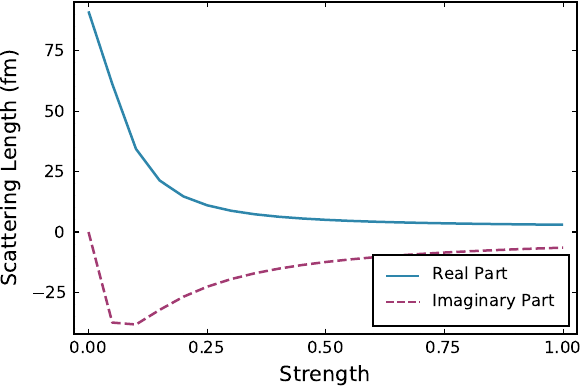}
    \label{subfig:scat_a}}
    \subfigure[]{\includegraphics[width=0.48\textwidth]{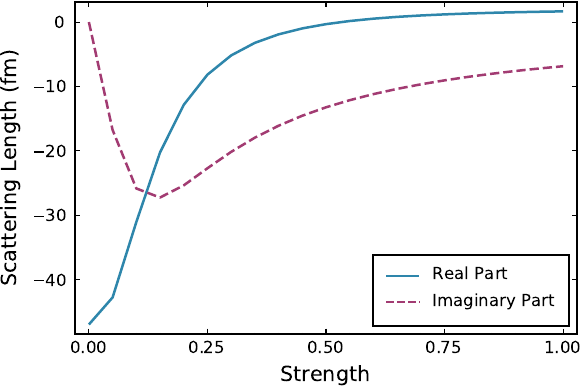}
    \label{subfig:scat_b}}
    \subfigure[]{\includegraphics[width=0.48\textwidth]{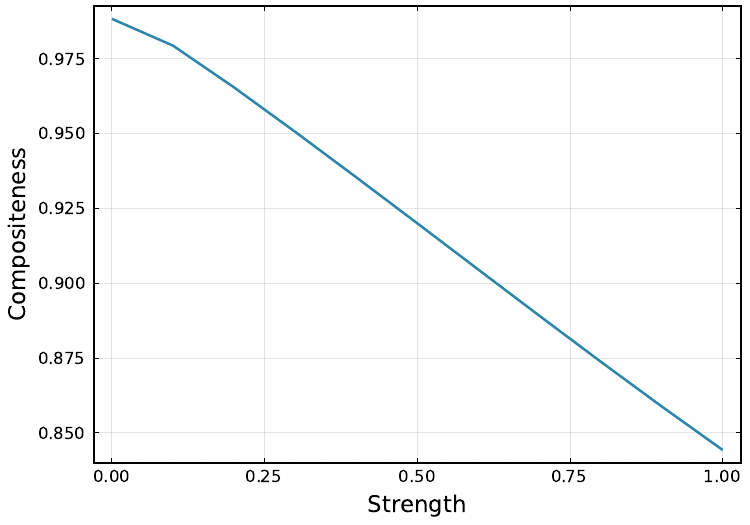}
    \label{subfig:comp_a}}
    \subfigure[]{\includegraphics[width=0.48\textwidth]{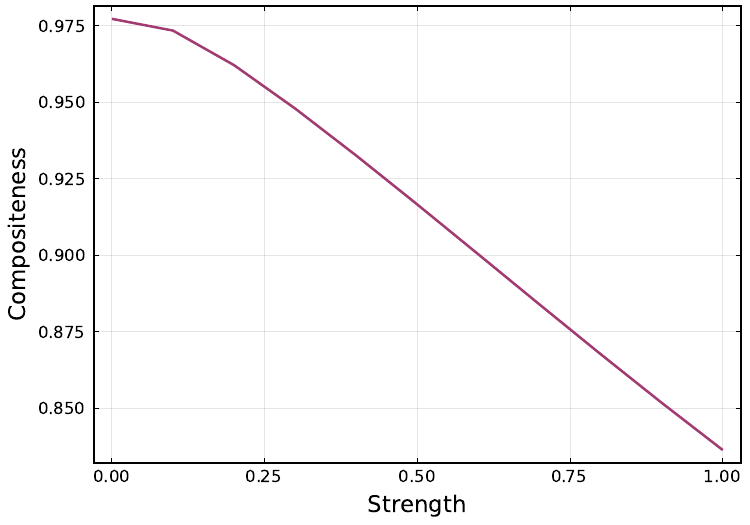}
    \label{subfig:comp_b}}
    
  \caption{Dependence of the scattering length and compositeness on the Pomeron exchange strength. (a,c) corresponds to the case where the scattering length is positive in the absence of Pomeron contribution, while (b,d) corresponds to the case where the scattering length is negative without Pomeron contribution.}
  \label{fig3}
\end{figure*}

\begin{figure*}[t]   
  \centering
    \subfigure[]{\includegraphics[width=0.8\textwidth]{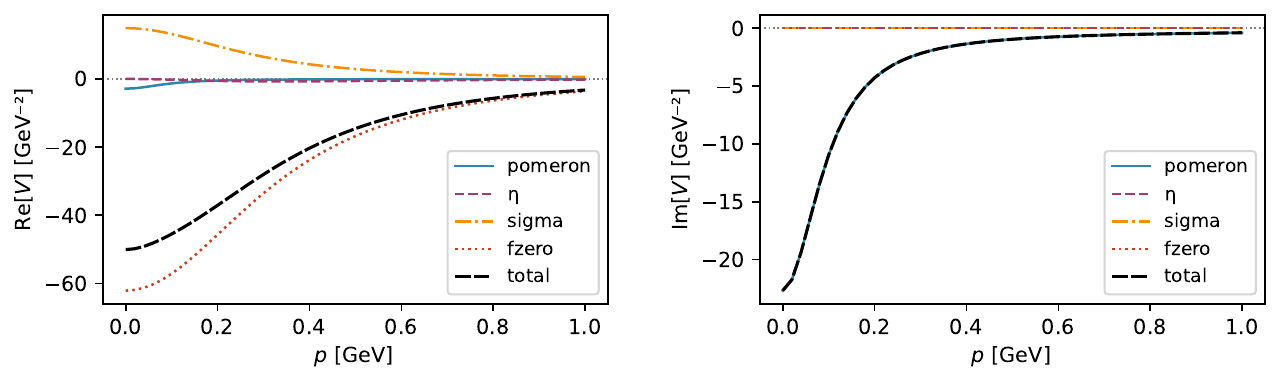}}
    \subfigure[]{\includegraphics[width=0.8\textwidth]{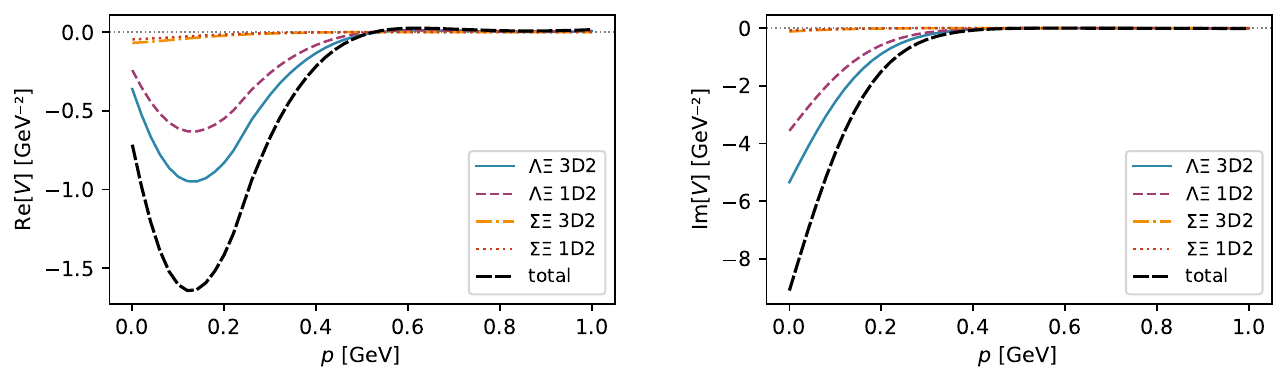}}
  \caption{The interaction potential of $N\Omega({^5}S{_2})$ in momentum space. (a) shows the leading-order contribution, and (b) corresponds to the box-loop diagram contribution. The value of $\Lambda_p$ is fixed as 0.2 $\mathrm{GeV}$. {The meaning of each curve?}} 
  \label{fig4}
\end{figure*}
\begin{figure*}[htbp!]  
  \centering
    \subfigure[]{\includegraphics[width=0.8\linewidth]{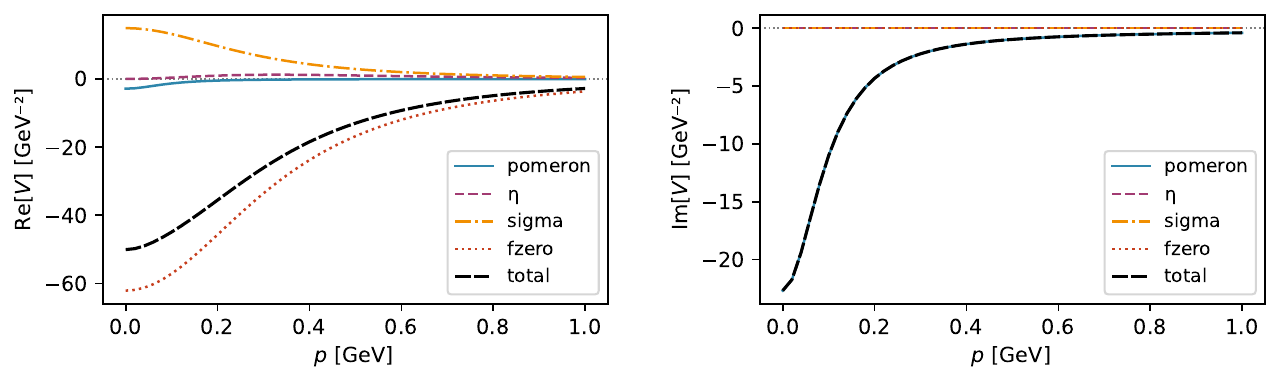}}
    \subfigure[]{\includegraphics[width=0.8\linewidth]{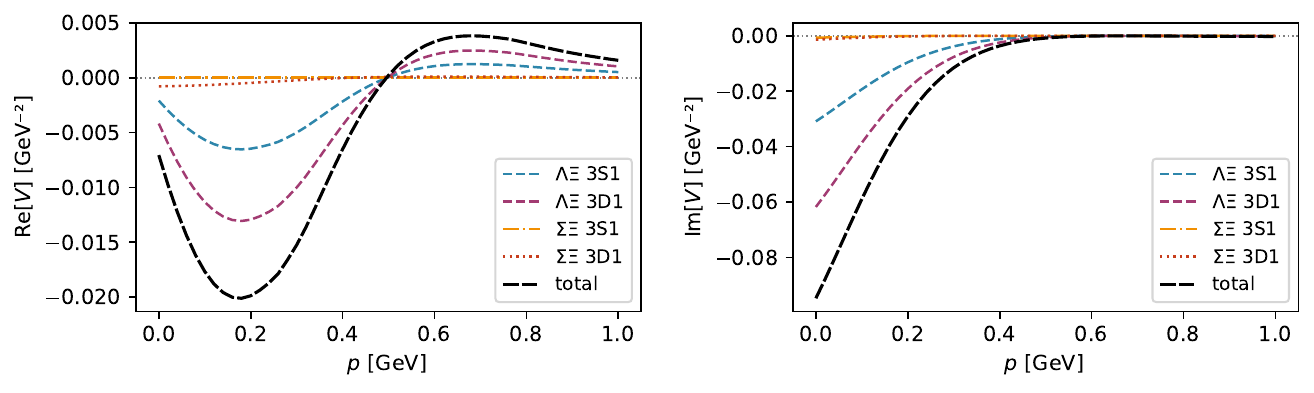}}
  \caption{The interaction potential of $N\Omega({^3}S{_1})$ in momentum space with the caption analogous to that of Fig.~\ref{fig4}. }
  \label{fig5}
\end{figure*}
\par
From the curves depicted in \cref{fig2}, we observe that as the cut-off $\Lambda_p$ increases, the interaction potential arising from the Pomeron exchange decreases more gradually with the increasing momentum. In the two-body center-of-mass frame, the momentum transfer $t$ is proportional to the momentum of the incident particle, and the Pomeron is required to satisfy the Regge condition $s \gg \left | t \right | $. For heavy hadrons the Regge condition can be approximately satisfied near threshold. But with the increase of $s$, the range of $t$ also increase, the Pomeron contribution will be  significantly suppressed at kinematics off forward. In the kinematic region of interest the Pomeron contribution turns out to decrease with $s$ going above threshold due to the form factor in Eq.~(\ref{FF-1}). The cut-off $\Lambda_p$-dependence in \cref{fig2} suggests that smaller values for the cut-off $\Lambda_p$ will lead to smaller contributions from the Pomeron exchange.

\par
\cref{Scattering Length,Effective Range} present the scattering length and effective range of the S-wave $N\Omega$ system for different cut-off values, where a cut-off of 0 corresponds to the case without the Pomeron contribution. We find that the real parts are in reasonable agreement with the experimental data and HALQCD calculations at small cut-off values around $\Lambda_p=0.15$ GeV, which is consistent with the requirement of a small $\Lambda_p$ imposed by the Regge condition discussed previously. From the experimental results of heavy-ion collisions, the spin-averaged effective range is larger than that of the quintet channel ${^5}S{_2}$ within uncertainties, which arises precisely because the effective range of the ${^3}S{_1}$ state is larger than that of the ${^5}S{_2}$ state. Notably, all our results are complex quantities, since both absorption effects and Pomeron exchange can contribute to the imaginary part of the interaction potential, which naturally introduces an imaginary component into the observable physical quantities.
In addition, we note that for the $S$-wave $N\Omega$ system, the scattering length decreases with the increasing cut-off. This arises because the Pomeron, as an effective description of soft gluon exchange, provides an extra attractive interaction that reduces the scattering length, as illustrated in \cref{subfig:scat_a}. 

To further investigate the influence of the Pomeron exchange strength on the properties of the pole, we perform a numerical test. Firstly, we only consider the scalar potential $V=V_\mathrm{s}$ and the variation of the scattering length with the strength of the attractive potential is shown in \cref{fig7}. We further consider the combined potential consisting of the scalar potential and the Pomeron potential, namely $V=V_{\mathrm{s}}+n*V_{\mathrm{P}}$. Here, two sets of $V_{\mathrm{s}}$ are adopted to yield positive and negative scattering lengths, respectively. The magnitude of the parameter $n$ is varied to characterize different strengths of the Pomeron exchange. The corresponding results are presented in \cref{fig3}. We find that the behavior differs notably when the scattering length is negative in the absence of the Pomeron contribution. Since the initial scattering length is negative, continuously strengthening the attractive interaction associated with the imaginary part drives the scattering length to cross zero and switch from negative to positive, as illustrated in \cref{subfig:scat_b}. This behavior is distinctly different from that of conventional interaction potentials containing only a real component, as illustrated by the first half of the curves in \cref{fig7}. Meanwhile, \cref{subfig:comp_a,subfig:comp_b} indicate that the Pomeron exchange renders the pole structure more compact.

\cref{fig4,fig5} display the interaction potential for the $N\Omega({^5}S{_2})$ and $N\Omega({^3}S{_1})$ systems with the same set of parameters in the momentum space, respectively. Notably, at leading order the interaction potentials for these two quantum states are almost the same, with the only distinction coming from the next-to-leading-order box loop diagrams. It is evident that the two systems should exhibit similar scattering behavior near the threshold, with the absorption effect giving rise to certain corrections.

We also search for poles of the S-wave $N\Omega$ scattering amplitude on the first Riemann sheet. Interestingly, removing the Pomeron-exchange contribution from our calculation, which corresponds to the $\Lambda_{p}=0$ case in \cref{Scattering Length} and \cref{fig6}, leaves the pole position essentially consistent with the HAL QCD result, but leads to a larger scattering length which is not consistent with the experimental measurement in the $^5S_2$ channel~\cite{Zhang:2025lfn}. It is therefore reasonable to include the contribution of the Pomeron in the $N\Omega({^5}S{_2})$ system. Our results support the existence of a quasi-bound state in $N\Omega({^5}S{_2})$ with a binding energy of $1.73~\mathrm{MeV}$ and a width of $2.9~\mathrm{MeV}$. An analysis of the box-loop diagram amplitudes indicates that $\Lambda \Xi$ is the dominant decay channel. A more detailed study of the decay properties has been presented in Ref.~\cite{Xiao:2020alj}, where the total decay width is estimated to range from a few hundred keV to slightly above 1 MeV. %The comparatively large width obtained in our calculation may indicate a possible double-counting issue between the Pomeron-exchange contribution and the absorption effect. 
Furthermore, the cutoff dependence of the pole position is illustrated in \cref{fig6}, and consistent with both experimental results and HALQCD calculations. 
In addition, a weaker quasi-bound state is identified in the $N\Omega({^3}S{_1})$ channel, with a binding energy of  $0.21~\mathrm{MeV}$ and a width of $2.08~\mathrm{MeV}$. This can be attributed to the weaker absorption effect in this channel, which leads to a reduced attractive interaction.

\section{Summary}\label{Sec-conclusion}

In this work we have investigated the $S$-wave $N\Omega$ interaction within the framework of the effective Lagrangian approach. The interaction potential is constructed via the meson exchange model and incorporated by the Pomeron exchange mechanism. By solving the L-S equation we obtain the full scattering amplitude. It shows that although the Pomeron exchange contribution is small, it is necessary to include it in order to match the experimental data and the HALQCD calculations.  
The scattering length and effective range have been calculated for various cutoff values, and the results at small cutoff are found to be consistent with the experimental data and HALQCD calculations. With the cutoff $\Lambda_{p}=0.15$ GeV, two quasi-bound states are obtained in the $N\Omega({^5}S{_2})$ and $N\Omega({^3}S{_1})$ channels, with pole positions at $2608.98-1.45i~\mathrm{MeV}$ and $2610.52-1.04i~\mathrm{MeV}$, respectively.

Our numerical results suggest that the Pomeron exchange plays an essential role in describing the near-threshold scattering behavior and bound-state structure of the $N\Omega({^5}S{_2})$ system. For the $N\Omega({^3}S{_1})$ channel, relevant experimental data and lattice QCD results are currently unavailable. Future experimental measurements on this channel will serve as an important basis for further verifying the contribution of the Pomeron exchange to the $N\Omega$ interaction.

\begin{acknowledgments}
Useful discussions with Xiao-Feng Luo and Meng-Lin Du are aknowledged. This work is partly supported by the National Natural Science Foundation of China with Grants Nos. 12235018,  12375073, and 12547105. 

\end{acknowledgments}

\bibliography{references.bib}
\end{document}